# Floquet-sideband-enhanced shortwave electrometry with Rydberg atoms

Jiaqi Yuan[1,*], Yunlong Xue[2,*], Yongjie Cheng[1,3], Ruidong He[1], Jiteng Sheng[4], Yanpeng Zhang[1], Zhengyang Bai[2,‡], Yu-Qiang Ma[2,5], Min Xiao[6], and Zhaoyang Zhang[1,†]

[1]*Key Laboratory for Physical Electronics and Devices of the Ministry of Education & Shaanxi Key Lab of Information Photonic Technique, School of Electronic Science and Engineering, Faculty of Electronic and Information Engineering, Xi'an Jiaotong University, Xi'an, 710049, China*

[2]*National Laboratory of Solid State Microstructures and School of Physics, Collaborative Innovation Center of Advanced Microstructures, Nanjing University, Nanjing, 210093, China*

[3]*National Key Laboratory of Metrology and Calibration, Beijing Institute of Radio Metrology and Measurement, Beijing, 100039, China*

[4]*Institute of Quantum Science and Precision Measurement, School of Physics, East China Normal University, Shanghai, 200062, China*

[5]*Hefei National Laboratory, Hefei, 230088, China*

[6]*Department of Physics, University of Arkansas, Fayetteville, Arkansas, 72701, USA*

[*]These authors contributed equally to this work.

Corresponding authors: [‡]zhybai@nju.edu.cn, [†]zhyzhang@xjtu.edu.cn

Rydberg atomic electric-field sensors, under the framework of optical excitation and readout, can overcome the size-to-wavelength constraint imposed by the Chu limit. However, their sensitivity for decametric-wavelength shortwave electric fields is substantially lower than that for microwave ones, stemming from the off-resonant nature of low-frequency signals with Rydberg transitions. Here, we demonstrate a high-sensitivity heterodyne shortwave sensor based on microwave-dressed Rydberg atoms, leveraging precisely modulated Floquet sidebands. Around local- and microwave-field-engineered Floquet sidebands, the steep response gradient, arising from the enhanced atom-shortwave interaction through additionally created coherent channels, induces pronounced amplification of the heterodyne intermediate-frequency signal. As a result, compared to the same atomic heterodyne setup without microwave modulation, such Floquet-sideband-enhanced shortwave measurement boosts the sensitivity by four orders of magnitude, yielding a sensitivity of $-122.7$ dBm·Hz$^{-1}$ for shortwave at 30 MHz. This work offers a potential route to high-sensitive portable shortwave receivers in radio astronomy, radar and long-distance communications.

**Teaser**

Microwave modulation dramatically enhances the sensitivity of shortwave electric-field measurement in Rydberg atomic vapors.

## INTRODUCTION

Shortwave electric fields, characterized by their decametric wavelengths, are essential for long-distance communications, over-the-horizon radar, meteorological monitoring, and even astronomical exploration. However, the Chu limit (*1*) dictates that the optimal shortwave detection via conventional metal antennas requires their physical dimensions to be comparable to the signal's wavelength. Such a requirement inevitably leads to large-scale receiving ends that are incompatible with portable applications, such as onboard and small satellite platforms. This is also true for radio observation in astronomy, which usually requires thousands of antenna units deployed in space beyond the ionosphere to avoid its shielding effect. Consequently, developing compact sub-wavelength antennas with high sensitivity for shortwave signals remains a persistent pursuit across diverse fields.

In recent years, Rydberg alkali atomic ensembles have emerged as a robust platform for electric-field metrology, offering SI-traceability and self-calibration based on fundamental atomic constants (*2-4*). This sensing paradigm operates by exciting atoms to high-lying Rydberg states (*5*), which are easily to be perturbed by the incident electric fields, thereby modifying the optical response of atoms (*6-10*). The dense distribution of Rydberg states provides numerous transitions that can resonantly couple to almost the entire microwave (MW) band (*11-14*). By selecting appropriate Rydberg transitions, high-sensitivity detection across various frequencies can be achieved within a single centimeter-sized vapor cell. This inherent flexibility allows Rydberg atomic sensors to circumvent the size-to-wavelength constraints imposed by the Chu limit, offering a promising pathway for the realization of miniaturized shortwave antennas.

Up to now, Rydberg-atom electric-field sensing has achieved significant success in ultra-sensitive MW metrology, leveraging strong resonant coupling between MW fields and two adjacent Rydberg states. This interaction manifests as Autler-Townes (AT) splitting (*6, 8*) in Rydberg electromagnetically induced transparency (EIT) spectrum (*15, 16*), where the splitting width scales linearly with MW strength. To further enhance the sensitivity, the atomic heterodyne technique (*17, 18*) exploits a robust local oscillator to generate coherent beat signals, enabling the simultaneous extraction of field amplitude, frequency, and phase. Generally, state-of-the-art Rydberg MW sensors reach sensitivity on the order of $nV\cdot cm^{-1}\cdot Hz^{-1/2}$ (*19-23*).

However, shortwave sensing performance is seriously compromised by the absence of resonant coupling to Rydberg transitions. Instead, these low-frequency signals primarily induce off-resonant Stark shifts (*24*) governed by the scalar polarizability, resulting in a diminished response. While electric-field detection between 10 kHz and 30 MHz has been demonstrated (*25, 26*), the sensitivity remains orders of magnitude inferior to that of resonant MW measurement. To address these limitations, various strategies have emerged, including the utilization of synthetic Floquet dimensions (*27*) and the exploitation of modulated Floquet sideband gaps (*28*). In general, the sensitivity reported for shortwave detection is mostly around $-75\ dBm\cdot Hz^{-1}$ (*26, 28, 29*) with Rydberg atomic vapor cell placed in free space. Recently, specialized architectures, such as planar waveguides (*30*) and subminiature shortwave resonators (*31*), have been developed to enhance the coupling between shortwave and atoms. Combining these structural innovations with well-designed sensing mechanisms could push the boundaries of low-frequency Rydberg electrometry toward unprecedented precision.

In this work, we demonstrate a high-sensitivity shortwave electric-field sensor of atomic heterodyne framework based on MW-dressed Rydberg atoms. A local shortwave field together with the MW field induces a strong AC Stark effect, generating a series of controllable Floquet sidebands across multiple Rydberg states, while the latter further facilitates the engineering of sidebands' relative positions. By matching the spacing of Floquet sidebands to the target shortwave frequency, there occurs additional coherent channels beyond the original EIT resonance. This significantly boosts the interaction between Rydberg atoms and shortwave for high-sensitivity detection. Upon the introduction of a weak target shortwave field, beat-note oscillations emerge within the transmission spectrum. These oscillatory signals are significantly amplified in the vicinity of the Floquet sidebands, with the assistance of the steep response gradient arising from the near resonant conditions. By exploiting this mechanism, we achieve a minimum detectable sensitivity of $-122.7\ dBm\cdot Hz^{-1}$ ($0.63\ \mu V\cdot cm^{-1}\cdot Hz^{-1/2}$), representing a significant improvement of over four orders of magnitude compared to the atomic heterodyne measurements in the same setup but without MW modulation. These findings underscore the pivotal role of Floquet engineering in controlling coherent processes in Rydberg atomic systems and provide a robust physical framework for high-precision shortwave electrometry.

Furthermore, the implementation within a centimeter-scale vapor cell offers a viable pathway toward high-performance and portable shortwave receivers.

## RESULTS

### Experimental scheme

We realize the high-sensitivity Rydberg-atom shortwave sensor with the experimental configuration illustrated in Fig. 1(a). Two laser fields are oppositely injected into a rubidium atomic vapor cell (at room temperature) in the collinear arrangement to drive a three-level ladder-type configuration [Fig. 1(b)], involving a highly-excited Rydberg state $49D_{5/2}$ (level $|r\rangle$). The weak probe laser (wavelength ~780 nm, frequency $\omega_p$, Rabi frequency $\Omega_p$) emitted by an external cavity diode laser excites the transition from $5S_{1/2}$ ($|g\rangle$) to $5P_{3/2}$ ($|e\rangle$). The coupling laser (~480 nm, $\omega_c$, $\Omega_c$) derived from a frequency-doubling laser system drives the transition $|e\rangle \rightarrow |r\rangle$.

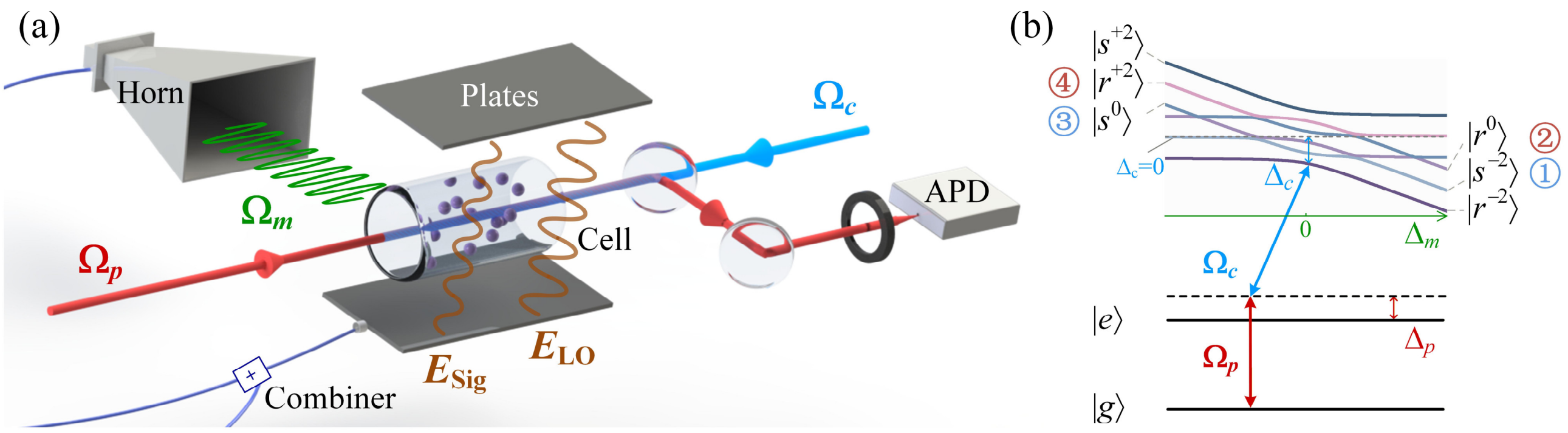

**Fig. 1. Scheme for shortwave measurement with Rydberg EIT-AT configuration.** (**a**) Experimental setup. The probe beam (red, with a wavelength of ~780 nm) overlaps with the counter-propagating coupling beam (blue, ~480 nm) inside the Rb atomic vapors (natural abundance) to prepare Rydberg atoms with the assistance of EIT. The local and signal shortwave fields are coupled to a pair of plates. APD: avalanche photodetector. (**b**) Energy spectrum as a function of the MW frequency detuning $\Delta_m$. The colored solid curves depict the six quasi-energy branches obtained by diagonalizing the Floquet effective Hamiltonian (see Supplementary Materials). For comparison, the bare atomic levels $|g\rangle$ and $|e\rangle$ are shown as black lines. The horizontal gray dotted line marks the position of $\Delta_c$=0. The Floquet quasi-energy sidebands are denoted by $|r^{-2}\rangle$, $|s^{-2}\rangle$, $|r^0\rangle$, $|s^0\rangle$, $|r^{+2}\rangle$, $|s^{+2}\rangle$. Detuning $\Delta_i$ ($i$=$m$, $p$ and $c$) is the difference between the frequency $\omega_i$ of field $\mathbf{E}_i$ and the energy gap of the two levels it connects.

A MW field $\Omega_m$ (with frequency being $\omega_m \approx 19.84$ GHz) drives the transition between two adjacent Rydberg states $49D_{5/2}$ ($|r\rangle$) and $48F_{7/2}$ ($|s\rangle$). Local ($\mathbf{E}_{LO}$, $\omega_{LO}$) and signal ($\mathbf{E}_{Sig}$, $\omega_{Sig}$) shortwave fields, featuring a frequency difference $\delta\omega = \omega_{Sig} - \omega_{LO}$, are together applied to a pair of electrode plates (7 cm spacing). The vapor cell is placed between the two plates, and the direction of the shortwave is perpendicular to that of the MW field. This dual-shortwave configuration allows for coherent heterodyne detection within the atomic ensemble. The combined effect of the local shortwave and MW fields induces a pronounced AC Stark effect on state $|s\rangle$, yielding tunable Floquet sidebands spanning multiple Rydberg levels, marked as $|r^{-2}\rangle$, $|s^{-2}\rangle$, $|r^{0}\rangle$, $|s^{0}\rangle$, $|r^{+2}\rangle$, $|s^{+2}\rangle$ in Fig. 1(b). In the meanwhile, the MW field enables the flexible and precise modulation on the positions of the formed sidebands.

**Theoretical model of Floquet modulation**

To describe the Floquet-engineered atomic system, we consider a four-level model consisting of the states ($|g\rangle$, $|e\rangle$, $|r\rangle$, $|s\rangle$). Under the rotating-wave approximation, the interaction Hamiltonian for optical, MW, and shortwave fields is given by

$$\begin{aligned}\hat{H}_I = &-\Delta_p\hat{\sigma}_{ee} - \left(\Delta_c+\Delta_p+\delta_r\right)\hat{\sigma}_{rr} - \left(\Delta_c+\Delta_p+\Delta_m+\delta_s\right)\hat{\sigma}_{ss} \\ &+\frac{1}{2}\left(\Omega_p\hat{\sigma}_{eg}+\Omega_c\hat{\sigma}_{re}+\Omega_m\hat{\sigma}_{sr}+\text{H.c.}\right)\end{aligned} \tag{1}$$

where $\Omega_p$, $\Omega_c$ and $\Omega_m$ denote the Rabi frequencies of the probe laser, coupling laser, and MW field, respectively, and $\Delta_p$, $\Delta_c$, and $\Delta_m$ represent their corresponding detunings. $\hat{\sigma}_{ij}=|i\rangle\langle j|(i, j=g, e, r, s)$ and $\delta_{r,s}$ denote the atomic transition operators and the energy shift of Rydberg states $|r\rangle$ and $|s\rangle$ caused by AC stark effect, respectively, and H.c. represent the Hermitian conjugate. This Hamiltonian accounts for the formation of Rydberg EIT as well as the AT splitting induced by the MW dressing field.

In the presence of the local shortwave electric field, the involved Rydberg energy levels experience AC Stark shifts. Under the weak-field approximation, these shifts can be expressed as $\delta_{r,s}(t)=-1/2\alpha_{r,s}E^2(t)$, where $\alpha_{r,s}$ denotes the scalar polarizability of Rydberg states $|r\rangle$ and $|s\rangle$, respectively, and $E(t)$ is the intensity of the total electric field. In the experiment, the composite field intensity is given by $E(t)=E_\text{d}+E_\text{LO}\cos(\omega_\text{LO}t)+E_\text{Sig}\cos[(\omega_\text{LO}+\delta\omega)t]$. Here, $E_\text{d}$ refers to the root-mean-square equivalent static component arising from unavoidable broadband background AC fields between the electrode plates. It is superposed with a strong local shortwave field $E_\text{LO}\cos(\omega_\text{LO}t)$ and a weak signal shortwave field $E_\text{Sig}\cos[(\omega_\text{LO}+\delta\omega)t]$. Given that $E_\text{d}$ is the background noise floor, its magnitude is significantly smaller than the local field, satisfying the field-strength hierarchy of $E_\text{LO} \gg E_\text{d}$ and $E_\text{LO} \gg E_\text{Sig}$.

Due to the quadratic AC Stark shift, the modulation naturally generates components at 0, $\omega_\text{LO}$, and $2\omega_\text{LO}$, which respectively mark the real energy level, the first-order Floquet sidebands, and the second-order Floquet sidebands. Since $\alpha_s \gg \alpha_r$ for $|r\rangle=|49D_{5/2}\rangle$ and $|s\rangle=|48F_{7/2}\rangle$, the modulation is dominated by the state $|s\rangle$ and the corresponding second-order terms at $\pm 2\omega_\text{LO}$ (see Supplementary Fig.

1). With the dressing effect from MW, the Rydberg levels split into a set of Floquet sidebands ($|r^{0,\pm 2}\rangle$ and $|s^{0,\pm 2}\rangle$), as shown in Fig. 1(b). The detailed Floquet theory and construction of sidebands in the MW-dressed Rydberg system accounting for Doppler thermal motion are provided in the Supplementary Materials.

**Demonstration of microwave-modulated Floquet sidebands**

To validate the Floquet quasi-energy model, we conduct both theoretical simulations and experimental measurements, as given in Fig. 2. Initially, the EIT spectrum exhibits well-defined AT splitting (under the interaction of MW), which is in excellent agreement with numerical simulations of the master equation derived from Hamiltonian (1) [see Figs. 2(a1), 2(a3) and 2(b1), 2(b3)]. Given the fact of relatively low strength ($E_{\text{LO}}$=8.3 mV·cm$^{-1}$) of the local field ($\omega_{\text{LO}}$=30 MHz), and of the small scalar polarizability from the state $|r\rangle$ ($\alpha_r \approx 30$ MHz·cm²/V², depending on the referred specific Rydberg level (*32, 33*)), the AC Stark shift is rather inconspicuous in the absence of MW dressing [Figs. 2(a2) and 2(b2)].

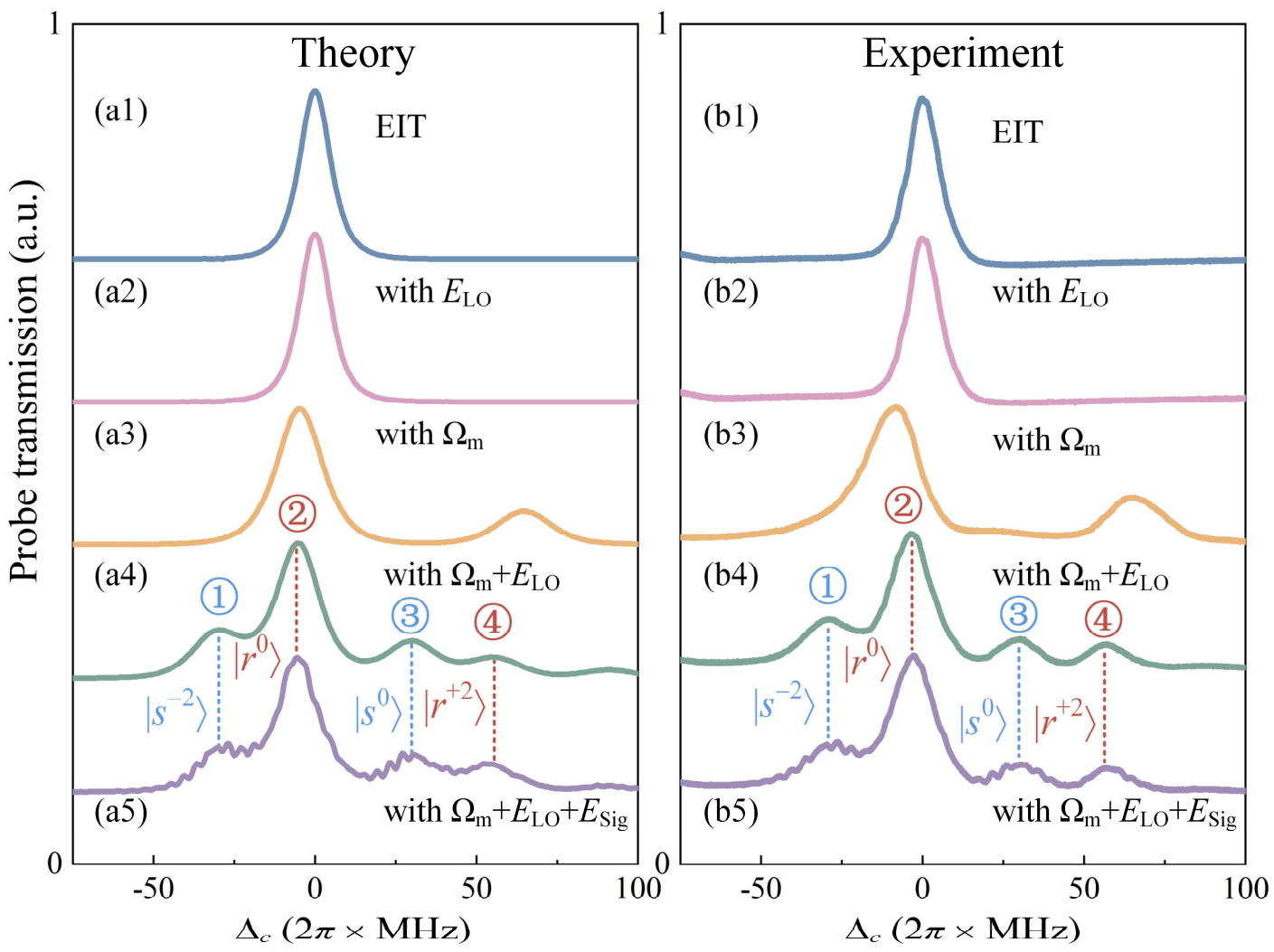

**Fig. 2. The transmitted probe spectra obtained by scanning $\Delta_c$ with different modulations in (a) theory and (b) experiment.** (**a1**) Rydberg EIT spectrum without electric-field modulation. (**a2**) Rydberg EIT spectrum with local shortwave field at $\omega_{\text{LO}}$=30 MHz. (**a3**) AT splitting caused by the detuned MW field ($\Delta_m$=−2$\pi$×80 MHz) with shortwave field off. (**a4**) The AT-splitting spectrum modulated by the same local shortwave field ($\mathbf{E}_{\text{LO}}$) as (a2). (**a5**) The spectrum under atomic heterodyne structure with the signal shortwave ($\mathbf{E}_{\text{Sig}}$) attending. Panels (**b1-b5**) show the experimental results, matching the simulated spectra in (a1-a5) from the Floquet theory. Peaks ①, ②, ③, and ④ correspond to states $|s^{-2}\rangle$, $|r^{0}\rangle$, $|s^{0}\rangle$, and $|r^{+2}\rangle$, respectively.

Notably, the simultaneous application of both the MW and local fields leads to the emergence of four distinct resonance peaks in the transmission spectrum. These peaks, labeled as ①–④, correspond to detunings of $\Delta_c$=−2$\pi$×30 MHz, −2$\pi$×3 MHz, 2$\pi$×30 MHz, and 2$\pi$×57 MHz, respectively, in Figs.

2(a4) and 2(a5). These resonances signify the emergence of Floquet sidebands originating from the MW-driven coupling between $|r\rangle$ and $|s\rangle=|48F_{7/2}\rangle$. Due to the large scalar polarizability of state $|s\rangle$ ($\alpha_s \approx 12665.4$ MHz·cm²/V²), a pronounced AC Stark effect occurs, inducing Floquet sidebands $|r^{\pm2,0}\rangle$ and $|s^{\pm2,0}\rangle$. Based on the energy-level analysis, peaks ①–④ are identified as $|s^{-2}\rangle$, $|r^{0}\rangle$, $|s^{0}\rangle$, and $|r^{+2}\rangle$. Notably, the frequency interval (~$2\pi\times60$ MHz) between peaks ①&③ and ②&④ matches the $2\omega_{LO}$ modulation frequency, confirming their origin as second-order Floquet signatures.

The introduction of a weak signal field $\mathbf{E}_{Sig}$ facilitates interference with $\mathbf{E}_{LO}$, thereby establishing an atomic heterodyne configuration. This interference manifests as a periodic intermediate-frequency (IF) signal in the probe transmission. As previously established (*17*), the amplitude of this IF signal is directly proportional to the product of the electric-field intensities ($E_{LO}\times E_{Sig}$). This identifies the local-field intensity as a gain factor that effectively “amplifies” the detection capability. As illustrated in Figs. 2(a5) and 2(b5), the beat-note oscillations at $\delta\omega$=1.7 kHz are predominantly enhanced in the near-resonant regions corresponding to peaks ① ($|s^{-2}\rangle$) and ③ ($|s^{0}\rangle$) with $E_{Sig}$=130 μV·cm$^{-1}$. Such enhancement arises from the fact that the response slope of the heterodyne signal amplitude with respect to the detuning $\Delta_m$ is significantly steepened around the optimal dressing conditions. A steeper slope in the transmission spectrum indicates a more pronounced modulation of the heterodyne signal amplitude per unit of shortwave electric field, thereby leading to an enhanced sensitivity for $\mathbf{E}_{Sig}$. Consequently, the resonant features of the $|s\rangle$-state Floquet sidebands render an enhanced transduction of the shortwave modulation into the probe transmission, thereby substantially improving the measurement sensitivity to shortwave fields.

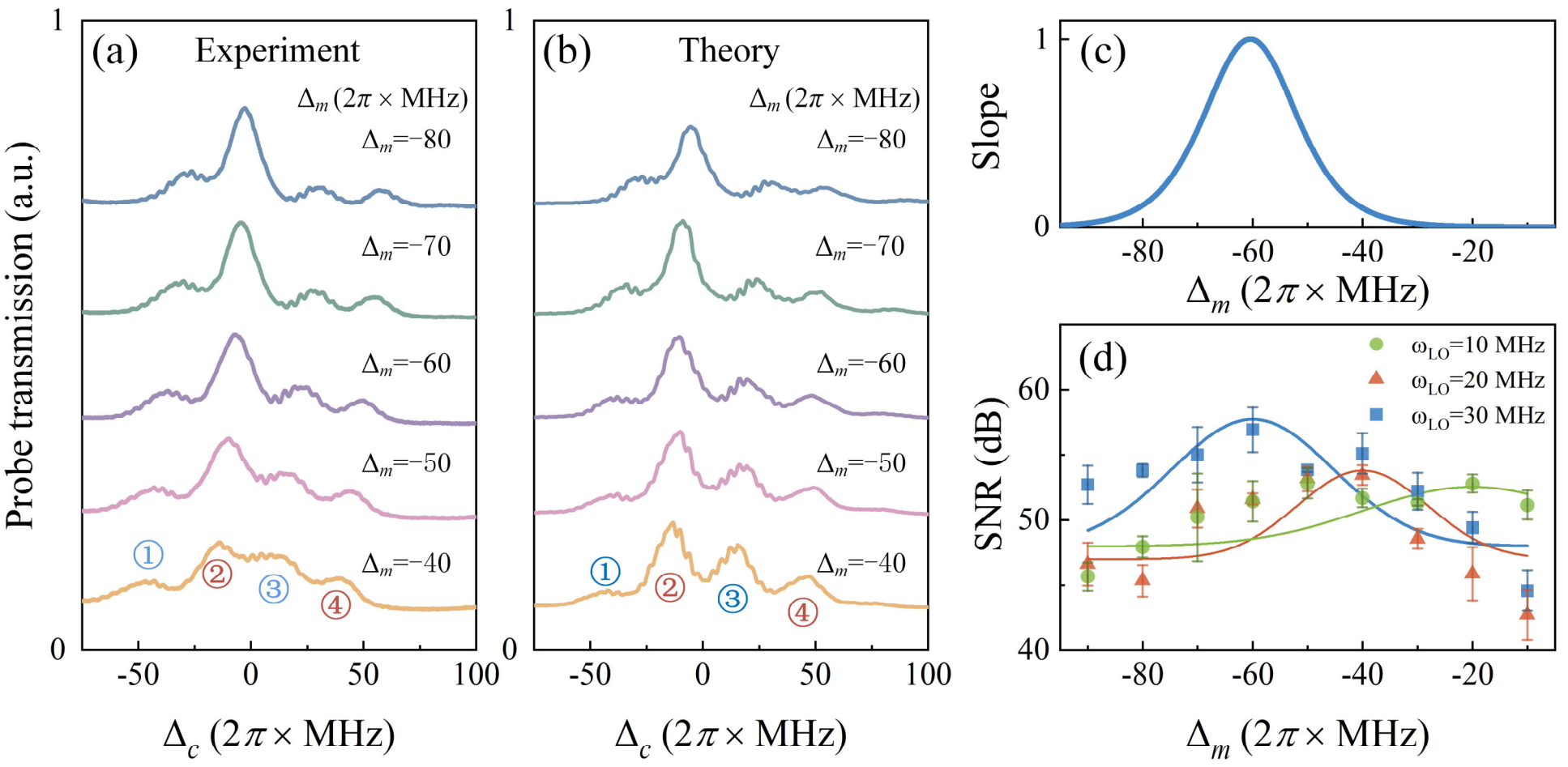

**Fig. 3. The sensitivity versus the MW detuning.** (**a**) The experimental and (**b**) theoretical evolutions of the probe transmission versus the coupling detuning $\Delta_c$ at different MW detuning $\Delta_m$ under the regime of atomic heterodyne.

The frequency of the shortwave is 30 MHz. (**c**) The relationship between response slope and MW detuning $\Delta_m$ at $\omega_{LO}$=30 MHz. (**d**) SNR of IF signals versus the MW detuning with fixed powers from two shortwave sources at different $\omega_{LO}$. The dots, triangles, and squares are the experimental results of $\omega_{LO}$=10, 20, and 30 MHz, respectively. SNR values are averaged from 3 measurements, with error bars showing the standard deviation. The corresponding solid curves, which are the mathematical fitting according to the experimental results, give a guide to the eye.

As illustrated by the effective energy spectrum in Fig. 1(b), the MW detuning governs the relative arrangements of the Floquet sidebands. To optimize the sensitivity of the shortwave measurement, we systematically investigate how this MW detuning influences the overall performance of the sensor. Figure 3(a) illustrates the experimental evolution of the Floquet-AT spectrum versus the MW detuning $\Delta_m$, with the signal frequency fixed at $\omega_{Sig}=\omega_{LO}$+1.7 kHz. The signal oscillations are primarily localized in state $|s\rangle$ and associated sidebands (peaks ① and ③), as the AC Stark effect originates predominantly from this state. As $\Delta_m$ varies from $-2\pi\times80$ MHz to $-2\pi\times40$ MHz, the oscillation amplitude first increases and then decreases, with the maximum occurring at $\Delta_m=-2\pi\times60$ MHz. These observations are well-supported by the corresponding simulations in Fig. 3(b).

The detection sensitivity is governed by the transduction gradient (i.e., the response slope) of the heterodyne amplitude with respect to the $\Delta_m$. Specifically, a steeper gradient yields a more pronounced amplitude for the periodic IF signal. By modulating the Floquet energy landscape and the setting of the sidebands, $\Delta_m$ directly shapes the response slope, thereby determining the system's ability to detect the shortwave signal. As illustrated in Fig. 3(c), the response slope exhibits a non-monotonic dependence on $\Delta_m$, which defines the regime for the optimal sensitivity. At a large detuning (e.g., $\Delta_m=-2\pi\times80$ MHz), the gradient is relatively small, resulting in limited measurement sensitivity. However, as $\Delta_m$ approaches the resonance condition, the slope increases significantly, indicating that the MW-induced enhancement of the shortwave measurement sensitivity is the most pronounced. This peak occurs at $\Delta_m\approx-2\omega_{LO}$, coinciding with the system's maximum response to the signal field. In contrast, for the case without the MW field, the same $\mathbf{E}_{LO}$ is too weak to induce a measurable AC Stark response in the Rydberg state $|r\rangle$ with a small scalar polarizability, leading to a flat (zero) response slope. This is consistent with the phenomena in Figs. 2(a2) and 2(b2).

Next, we examine the evolution of the Signal-to-Noise Ratio (SNR) as a function of the MW detuning $\Delta_m$ with $\Delta_p$=0 and $\Delta_c=2\pi\times15$ MHz (corresponding to peak ③). The blue curve in Fig. 3(d) depicts the case with the signal-field frequency and power maintained at 30.0017 MHz and −65 dBm, respectively. Here, the SNR is averaged from the FFT spectra of IF signals acquired over three repeated measurements, with error bars denoting standard deviation. As $\Delta_m$ varies from $-2\pi\times90$ MHz to $-2\pi\times10$ MHz, the SNR exhibits a similar evolution to that of the slope in Fig. 3(c), initially rising to the maximum before subsequently declining. This indicates the existence of an optimal $\Delta_m$ that yields the best SNR (thus the highest sensitivity) for a given signal-field frequency. Similar trends are observed for $\omega_{LO}$=10 MHz and 20 MHz (pink and orange curves), confirming the robustness of this sensing mechanism across different frequency regimes, but the $\Delta_m$ for the highest SNR relies on $\omega_{LO}$.

As $\omega_{LO}$ increases from 10 MHz to 30 MHz, the corresponding optimal MW detuning $\Delta_m$ shifts from approximately $-2\pi\times20$ MHz to $-2\pi\times60$ MHz. This shift originates from the fact that a higher $\omega_{LO}$ broadens the energy spacing between Floquet sidebands, necessitating a proportionally larger $\Delta_m$ to maintain the optimal resonance matching for maximum sensitivity. Consequently, a larger $\Delta_m$ is required to achieve optimal sensitivity at higher $\omega_{LO}$. In principle, the established system exhibits consistent sensing performance across various signal frequencies. However, the measured SNR for the 10 MHz signal remains relatively low. This is attributed to the inevitable screening effect for low-frequency electric fields induced by free charges (originating from atomic ionization) accumulating on the inner surface of the vapor cell (*34-37*). The occurred shielding effect can be effectively suppressed by applying a vapor cell of sapphire or by placing the electrode plates inside the cell(*38, 39*).

**Floquet-sideband-enhanced shortwave electrometry**

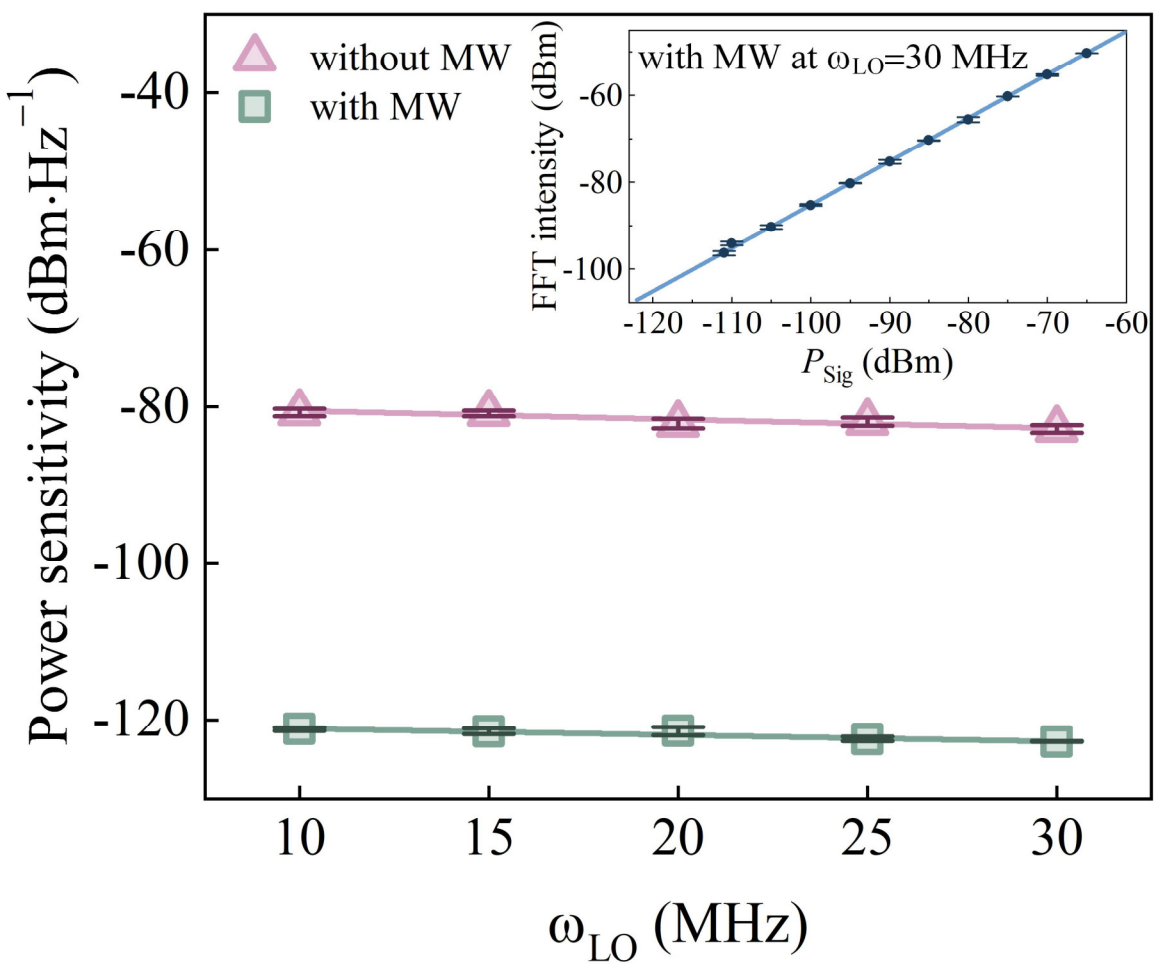

**Fig. 4. Measurement of the power sensitivity for signals of different frequency.** The measured power sensitivity at different shortwave frequencies without (dots) and with (squares) MW modulation in the same setup of atomic heterodyne structure. The corresponding solid curves are the mathematical fitting according to the observations. Inset: FFT intensities versus $P_{\mathrm{Sig}}$ for the situation of MW modulation at $\omega_{LO}$=30 MHz, where the triangles represent the experimental results and the solid line denotes the linear fitting curve. The FFT intensities are averaged from 3 measurements, with error bars showing the standard deviation.

Finally, to evaluate the Floquet-sideband enhancement on shortwave measurement, we characterized the variation of the power sensitivity over the signal frequency ($\omega_{LO}$ ranges from 10~30 MHz with a step of 5 MHz) with and without MW modulation, respectively. As shown by the squares in Fig. 4, by introducing MW modulation and optimizing parameters such as $\Omega_m$ and $E_{LO}$ for each frequency point, the established sensing system exhibits an optimal (minimum detectable) power sensitivity of −122.7 dBm·Hz$^{-1}$ (−121.2 dBm·Hz$^{-1}$) at $\omega_{LO}$=30 MHz (10 MHz), corresponding to an intensity sensitivity of 0.63 μV·cm$^{-1}$·Hz$^{-1/2}$ (0.75 μV·cm$^{-1}$·Hz$^{-1/2}$). The inset presents the variation of measured FFT intensity (triangles) with the signal-field power sensed by the beam $P_{Sig}$ under MW modulation at $\omega_{LO}$=30 MHz. Clearly, the FFT intensity exhibits almost a linear dependence on $P_{Sig}$. Compared to the results without MW modulation (dots), the scheme with MW achieves an improvement of ~40 dB in power sensitivity. The linear correspondence between FFT intensity and $P_{Sig}$ at other frequencies in cases with and without MW can be found in Supplementary Fig. 4.

**DISCUSSION**

In summary, this work proposes and implements an MW-modulated approach to enhance the sensing sensitivity of shortwave signals with the atomic heterodyne structure. The Floquet sidebands are engineered via the combined action of a MW field near-resonant with Rydberg energy levels and a strong local shortwave field. The scheme achieves a high sensitivity of −122.7 dBm·Hz$^{-1}$, corresponding to the intensity sensitivity of 0.63 μV·cm$^{-1}$·Hz$^{-1/2}$. This represents a 4-order of magnitude improvement over conventional off-resonant atomic heterodyne technique, significantly boosting the shortwave measurement capability of Rydberg atoms. There is no doubt that such a highly sensitive mechanism relying on centimeter-scale atomic vapor cell will make great contributions to the development of portable shortwave receivers. Promisingly, this sensing method could apply to other low-frequency electrical signals that cannot resonantly couple with Rydberg levels. The current work exhibits tremendous potentials in quantum metrology, long-distance communications, and other applications involving low-frequency electrical signals. For example, the detection of signals from 0.1 MHz to 30 MHz is of high scientific value in radio astronomy. However, conventional antennas face challenges in covering such a wide frequency window while maintaining sufficient gain. The demonstrated high-sensitivity Rydberg shortwave sensor, which supports portable receivers operating across this broad frequency range, provides an alternative pathway toward to overcome this challenge.

As a proof-of-concept demonstration, the atomic vapor cell for receiving shortwave fields is centimeter-scale. However, the adopted laser, local shortwave and microwave systems for manipulating Rydberg atoms are much larger that the vapor cell, which hinders immediate integration into truly compact and portable receivers. Fortunately, the development of chip-scale microwave sources (*40*) and fiber-coupled compact frequency-stabilized diode lasers (*41-43*), could help to construct a miniaturized driving system for the proposed sensing strategy.

## MATERIALS AND METHODS

### Experimental settings

The frequencies of probe and coupling field are locked to the saturation absorption and reference EIT spectra, respectively. The reference EIT is produced by two beams from the same sources as $\Omega_p$ and $\Omega_c$. The frequencies of both laser beams are further controlled by individual frequency-shift systems constructed with acoustic optical modulators. The radii of $\Omega_p$ and $\Omega_c$ are 0.25 mm and 0.4 mm. The powers of $\Omega_p$ and $\Omega_c$ beams are 30 μW and 80 mW, corresponding to Rabi frequencies of $\Omega_p$=2$\pi$×10 MHz and $\Omega_c$=2$\pi$×2.4 MHz, respectively. The output probe field is reflected by a dichroic mirror after the Rb cell and focused into an avalanche photodetector to capture the transmission spectrum. The output of the photodetector is simultaneously connected to the oscilloscope and the spectrum analyzer to achieve synchronous observation and acquisition of time-domain and frequency-domain signals. The spectrum analyzer is set to a resolution bandwidth of 1 Hz and a video bandwidth of 3 Hz. The two shortwave fields are combined together with a RF combiner. Then they are sent to the electrode plates to radiate the vapor cell with a length of 10 cm. The calibration of the shortwave-field loss from the cables and electrode plates is provided in the Supplementary Materials.

## Acknowledgements

**Funding:**
National Natural Science Foundation of China (62475209, 12274131, 12347102)
Qinchuangyuan “Scientist+Engineer” Team Construction of Shaanxi Province (2024QCY-KXJ-178)
Key Research and Development Program of Shaanxi Province (2025PT-ZCK-49)
Natural Science Foundation of Jiangsu Province under Grant (BK20233001)
Innovation Program for Quantum Science and Technology under Grant (2024ZD0300101)

**Author contributions:**
Conceptualization: Z. Z.
Experiment: J. Y., Y. C., R. H.
Theoretical support: Y. X., Z. B.
Interpretation: J. S., Y. Z., M. X., Y. M.
Supervision: Z. Z., Z. B.
Writing—original draft: J. Y., Y. X.
Writing—review & editing: Z. Z., Z. B.

**Competing interests:** The authors declare no competing interests.

**Data, code, and materials availability:** All data required to evaluate and reproduce the results within this paper are available from the corresponding authors upon reasonable request.